\documentclass[11pt]{article}

\usepackage{amsmath}
\usepackage{comment}

\numberwithin{equation}{section}

\begin{document}


\title{\bf  Geodesic multiplication as a tool for\\ 
classical and quantum gravity\footnote{Trans. Tallinn Tech. Univ. {\bf733} (1992), 33-42.}}

\author{\Large Piret Kuusk and Eugen Paal} 

\date{}
\maketitle

\thispagestyle{empty}

\begin{abstract}
Algebraic systems called the local geodesic loops and their tangent Akivis algebras are considered. 
Their possible role in theory of gravity is considered. Quantum conditions for the infinitesimal quantum events 
are proposed.
\end{abstract}



\section{Introduction}

Spacetime and event are two fundamental concepts of relativistic physics.
According to theory of general relativity, totality of all events or the
classical (pre-quantum) spacetime can be represented 
by a ($1+3$)-dimensional Lorentzian manifold. 
Evolution of the Lorentzian metric tensor is prescribed 
by the Einstein equations, constraining the Einstein and 
energy-momentum tensors of spacetime to be proportional. 
On the classical level, gravitation reveals itself through 
curvature of spacetime. In this sense the general 
relativity may be proclaimed to be a geometrical theory of 
the gravitational field

There have been several attempts to quantize gravity. 
In fact, quantization of the graviatational field is one 
of the most intriguing problems in theoretical physics. 
In a sense, quantization is first of all an algebraic 
method for one looks for the quantum observables  which 
must imitate the algebraic properties of the classical ones. 
So every algebraic aspect of the classical spacetime must 
be thoroughly taken into account as well.

In this paper we outline an idea of the geodesic quantization, 
based on construction of the local geodesic multiplication ($GM$) 
of spacetime points (Sec.~2). For the Minkowski spacetime of 
special relativity, the $GM$ in fact coincides with the common 
vector addition rule. Hence, in this (Minkowskian) case the $GM$ 
turns out to be a globally defined commutative and associative 
operation. The Abelian property of the $GM$ is in fact due to 
the globally vanishing torsion and curvature of the Minkowski space.

Geodesic multiplication \cite{1,2,3} on manifolds with an affine 
connection is mathematically rather elaborated but is still 
not widely known for physicists. Nevertheless, it deserves attention as well. 

Via the $GM$ on can treat spacetime algebraically, as a 
collection of the algebraic systems called the local geodesic loops. 
In general relativity, the $GM$ acquires dynamical meaning. 
Nonassociativity of $GM$ is an algebraic manifestation 
of the curvature of spacetime (Sec.~3). Tangent spaces at  
spacetime points turn out to be binary-ternary algebras called 
the Akivis algebras. Based on this fact, the geodesic quantum 
conditions are proposed for the infinitesimal quantum events (Sec.~4).

\section{Local geodesic multiplication}

At first let us introduce some basic algebraic notions.

A quasigroup \cite{4,5} is a set $G$ with a binary operation (which we 
denote by juxtaposition) which has the following property: 
in equation $gh = k$, knowledge of any two elements specifies the
third one uniquely. A quasigroup with a unit element is called a loop \cite{4,5}. 

For a fixed point $e$ of spacetime $M$, choose a tangent vector $X$ from
the tangent space $T_{e}(M)$ of $M$ at $e$. Consider a local path
$t \mapsto g(t;X)$ in $M$ through point $e$ with the tangent
vector $X$ at $e$:
\begin{equation}
g^{i}(0;X) = e^{i},\quad
\partial_{t}g^{i}(0;X) = X^{i}. \label{8}
\end{equation}
It is well known that this path is the unique local geodesic path
through $e$ in direction $X$ iff the following differential
equation holds:
\begin{equation}
\partial^{2}_{t}g^{i} + \Gamma^{i}_{lm}\partial_{t}g^{l} \partial_{t}g^{m} = 0, \label{9}
\end{equation}
where $\Gamma^{i}_{lm}$ are the affine connection coefficients.

The exponential mapping $X \mapsto g : = \mbox{Exp}_{e}(X) :=
g(1;X)$ at $e$ is known \cite{6} to be a local diffeomorphism of a
suitable neighbourhood of origin of $T_{e}(M)$ onto the corresponding
(normal) neighbourhood of $e\in M$.
Note that this property allows us treat the tangent vectors of
the spacetime as infinitesimal events. We can also say that every event
from the normal neighbourhood of $e$ can be generated via the
exponential mapping by the corresponding tangent vector from
$T_{e}(M)$.

The local geodesic loop at $e$ can be constructed in such a
neighbourhood $M_{e}$ of $e$, where all required exponential mappings are well defined
local diffeomorphisms. Choose in $M_{e}$ another local geodesic arc
$ h(s;Y)$ through the point $e$ with the direction
$Y \in T_{e}(M)$. To perform the parallel transport of $X \in T_{e}(M)$
along this geodesic, we must solve the linear Cauchy problem
\begin{equation}
\partial_{s}X^{\prime i} + \Gamma^{i}_{ml} \partial_{s}h^{l} X^{\prime m} = 0,
\quad X^{\prime}(0) = X.    \label{10}
\end{equation}
Performing the parallel transport of $X \in T_{e}(M)$, we obtain at
$h := \mbox{Exp}_{e}(Y)$ the tangent vector $X^{\prime} := X^{\prime}(1)$ in
$T_{h}(M)$. Now, draw the local geodesic arc through $h$ in the direction
$X^{\prime}$, and mark point $\mbox{Exp}_{h}(X^{\prime})$ on it.
This point is called the product of $g$ and $h$, and it will be denoted as
$gh$. Explicitly, the multiplication formula
reads \cite{1}
\begin{equation}
gh= (\mbox{Exp}_{h} \circ \tau^{e}_{h} \circ \mbox{Exp}^{-1}_{e})g,
\label{11}
\end{equation}
where $\tau^{e}_{h}: T_{e}(M) \rightarrow T_{h}(M)$ denotes the parallel
transport mapping of the tangent vectors from $T_{e}(M)$ into $T_{h}(M)$
along the unique local geodesic arc joining points $e$ and
$h: \tau^{e}_{h}(X) = X^{\prime}$.
The neighbourhood $M_{e}$ of $e$ with multiplication rule (\ref{11}) is  a local differentiable loop
\cite{1,2,3} denoted henceforth by $M_{e}$ as well. The unit element of
$M_{e}$ is $e$, and the local geodesic paths through the unit element $e$ are the
one--parameter subgroups of $M_{e}$. One can also say that the local
geodesic multiplication is monoassociative:
\begin{equation}
gg \cdot g = g \cdot gg,   \quad \forall g\in M_{e}.
\end{equation}
Note that the crucial part of the construction lies on Cauchy
problems  (\ref{8}),  (\ref{9})  and (\ref{10}), on existence and
uniqueness of their solutions, and also on the local diffeomorphism
property of the exponential mapping \cite{6}.

We can repeat the above construction and attach a local geodesic
loop to all reasonable points of  the spacetime. The patching conditions
for the local geodesic loops attached to different points of a manifold
with affine connection have been described by L.V.~Sabinin \cite{7}.

One can easily check that all geodesic loops of the Minkowski spacetime
are the Abelian groups. In this particular case, the geodesic multiplication
can be found from the common vector addition rule. The Abelian
property manifests algebraically the fact that the affine spaces are
globally torsionless and flat.

\section{Akivis algebras}

Generally speaking, the geodesic loops need not be commutative
and associative. There may exist such a triple of points
$g$, $h$, $k$ in $M_{e}$ that
\begin{equation}
gh \not= hg, \quad gh \cdot k \not= g \cdot hk.
\end{equation}
Let us choose in $M_{e}$ the local coordinates where $e^{i} = 0$
for all $i$. Deviation of $M_{e}$ from commutativity
and associativity can be measured \cite{3} by the structure
constants $c^{i}_{lm}$ and $A^{i}_{lmn}$ defined by
\begin{align}
(gh)^{i} - (hg)^{i} &= C^{i}_{lm} g^{l} h^{m} + \cdots,\\
(gh \cdot k)^{i} - (g \cdot hk)^{i} &= A^{i}_{lmn} g^{l} h^{m} k^{n} +\cdots,
\end{align}
where dots mean the higher order terms. It turns out that
non-commutativity and non-associativity of the local geodesic loops
are intimately related to torsion and curvature of spacetime.
Denote the torsion and curvature tensors as 
$S^{i}_{lm}$ and $R^{i}_{lmn}$, respectively. The direct
computations \cite{3} show that
\begin{eqnarray}
C^{i}_{lm} &=& 2S^{i}_{lm}(e),
\label{24}\\
A^{i}_{lmn} &=& R^ {i}_{lmn}(e) -
\bigtriangledown_{n} S^{i}_{lm}(e), \label{25}
\end{eqnarray}
where $\bigtriangledown_{n}$ denotes the covariant differentiation
operator.

We can now introduce \cite{8,9} the tangent algebra $A_{e}$ of 
$M_{e}$ similarly to the tangent (Lie) algebra of a Lie group. 
Geometrically, the tangent algebra $A_{e}$ 
coincides with the tangent space $T_{e}(M)$ of $M_{e}$ at $e$.
The product $[X,Y]$ of $X,Y \in A_{e}$ is defined in $A_{e}$ by
\begin{equation}
[X,Y]^{i} := C^{i}_{lm} X^{l} Y^{m} = -[Y,X]^{i}.
\label{31}
\end{equation}
We can equip $A_{e}$ with a ternary operation as well \cite{3,10}.
For a triple $X,Y,Z \in A_{e}$, define their triple product $(X,Y,Z)$ in
$A_{e}$ by
\begin{equation}
(X,Y,Z)^{i} := A^{i}_{lmn} X^{l} Y^{m} Z^{n}.
\label{32}
\end{equation}
The tangent algebra $A_{e}$ is thus a binary--ternary algebra, and it
need not be a Lie algebra. In other words, there may be a triple
$X,Y,Z \in A_{e}$, such that the Jacobi identity fails in $A_{e}$:
\begin{equation}
J(X,Y,Z) := [[X,Y],Z] + [[Y,Z],X] + [[Z,X],Y] \not= 0. \label{33}
\end{equation}
Instead, for all $X,Y,Z$  in $A_{e}$, we have \cite{3} a more general
identity 
\begin{equation}
J(X,Y,Z) = (X,Y,Z) + (Y,Z,X) + (Z,X,Y) -  (X,Z,Y) - (Z,Y,X) - (Y,X,Z) \label{34}
\end{equation}
called the  Akivis identity. The binary-ternary algebra $A_{e}$ is
hence called the  Akivis algebra.

{\it Comment}. It is well known that the tangent algebras of the local Lie groups are the Lie algebras \cite{11}. 
Non--Lie Akivis algebras appeared first as the tangent algebras of the local analytic loops \cite{10}. 
The tangent algebras of the local analytic Moufang (Bol) loops turn out to be the Mal'tsev (Bol) algebras \cite{8} (\cite{12,13,14}). These cases are quite remarkable for the following reason (generalized converse third Lie theorem): every real finite--dimensional Mal'tsev (Bol) algebra is the tangent algebra of some analytic Moufang (local Bol) loop \cite{15,16,17} (\cite{12,13,14}). The converse third Lie theorem for the general Akivis algebras has
been discussed in \cite{10,18}.

\section{Quantum conditions}

One can construct the Akivis algebras via non--associative algebras as well.
If we denote multiplication of a non--associative algebra by
juxtaposition, then its commutator--associator algebra ($CA$--algebra) is
the one with the following commutator and associator brackets:
\begin{equation}
[x,y] := xy - yx,\quad  (x,y,z) := xy \cdot z - x \cdot yz. \label{35}
\end{equation}
Anti--commutativity of the commutator bracketing is obvious, and the Akivis
identity can be checked by direct calculations. The original
non--associative algebra can be said to be the enveloping algebra of the
corresponding $CA$--algebra.

{\it Comment}. It is well known that the $CA$--algebras of associative algebras are the Lie algebras. $CA$--algebras of alternative algebras are the Mal'tsev algebras \cite{8}. $CA$--algebras of right--alternative algebras turn out to be the Bol algebras \cite{13,20}. The problem of  imbedding of non--Lie Akivis algebras into non--associative (enveloping) algebras (generalization of the Birkhoff--Witt theorem) is posed in \cite{10} and has not been solved yet\footnote{This problem has been solved by I.~P.~Shestakov in Dokl. Akad. Nauk {\bf368} (1999), 21--23.}.

We can exploit the above--presented property of the $CA$--algebras and propose the quantum conditions
for the infinitesimal quantum events, called the geodesic quantum conditions.

In a sense, every quantization is a representation of classical observables:
algebraic properties of quantum observables are believed to imitate
algebraic properties of the classical ones. Otherwise we are confronted
with anomaly (quantum mechanical symmetry breaking) \cite{21,22,23}.
For example, in canonical quantization, the canonical algebraic
structure of observables is required to be preserved. Likewise,
we can try to preserve the algebraic structure of the classical
infinitesimal events for the quantum ones as well.

Geodesic quantization must be a correspondence between the classical and
quantum events. For an infinitesimal event $X$ (tangent vector), let us
denote the corresponding quantum (infinitesimal) event as $Q_{X}$. If we
believe that the infinitesimal quantum events preserve
the structure of geodesic
Akivis algebras, the geodesic quantum conditions read 
\begin{align}
[Q_{X},Q_{Y}] &:= Q_{X}Q_{Y} - Q_{Y}Q_{X} = q Q_{[X,Y]}, \label{36}\\
(Q_{X},Q_{Y},Q_{Z}) &:= Q_{X}Q_{Y} \cdot Q_{Z} - Q_{X} \cdot Q_{Y}Q_{Z} =q^{2} Q_{(X,Y,Z)}. \label{37}
\end{align}
Here, the quantization constant (geodesic quantum deformation parameter)
is denoted as $q$, and the classical infinitesimal events $X,Y,Z$ must
belong to the same geodesic Akivis algebra. For $X,Y,Z$ from $A_{e}$, the
brackets in (\ref{36}), (\ref{37}) are defined by (\ref{31}),(\ref{32}).

Conditions (\ref{36}), (\ref{37}) mean that we seek for the enveloping
non--associative algebra of the geodesic Akivis algebra. In this case,
the binary and ternary multiplications of the infinitesimal events are
concealed, respectively, into commutator and associator of the
corresponding enveloping non--associative algebra. Non--associativity
of the quantum events is the price we must pay for the concealing, but
we get in fact rid of the ternary structure of the Akivis algebra,
which seems to be quite a beneficial and sensible compensation. Recall
that in the canonical quantization, the Poisson algebra multiplication of
the classical observables is concealed into associative multiplication of
the enveloping associative algebra of the quantum observables.

We can also assume that the geodesic quantization rule $X \rightarrow Q_{X}$
is linear. This means that for $X = X^{i}\partial_{i}$ the corresponding
quantum event $Q_{X}$ reads
\begin{equation}
Q_{X} := X^{i}Q_{i} \label{38}
\end{equation}
and (\ref{36}), (\ref{37}) read
\begin{align}
[Q_{l},Q_{m}] 
&:= Q_{l}Q_{m} - Q_{m}Q_{l} = q S^{i}_{lm}Q_{i}, 
\label{39}\\
(Q_{l},Q_{m},Q_{n}) 
&:= Q_{l}Q_{m} \cdot Q_{n} - Q_{l} \cdot Q_{m}Q_{n} = q^{2}(R^{i}_{lmn} - \nabla_{n}S^{i}_{lm}) Q_{i}. \label{40}
\end{align}
Here, the torsion and the curvature tensors must be valued at the point
where the tangent vectors (infinitesimal events) $\partial_{i}$ are
taken at.

Consistency of the geodesic quantum conditions (\ref{36})--(\ref{40}) with general
relativity (Einstein equations), and also compatibility (patching)
conditions of the quantizations at different spacetime points
must be inquired. The construction of observables
by means of the non--associative quantum events will be a crucial
problem as well.

Finally, let us note that in the early days of quantum mechanics
Jordan, von Neumann and Wigner \cite{24,25,26,27} tried to describe the quantum
observables in terms of  commutative but non-associative
(power-associative) algebras, nowadays called the
commutative  Jordan algebras. In \cite{28,30}, non-associativity
was suggested to be related with the elementary length.
Thorough historical review and discussion about the physical
meaning and evolution of non-associative structures in physics
is presented in \cite{30,31}.

\section*{Acknowledgement}

The present paper is dedicated to the memory of Prof. J\"uri Nuut, who wrote the first popular book in Estonian on the Einstein theory of relativity \cite{32}, read the first lecture course on mathematical foundations of the relativity theory in Tartu University (1932/33) and stood at the beginning of the research work on non-Euclidean geometry and cosmology in Estonia \cite{33}.

We would like to thank Prof. J.~Lukierski for invitation to Wroc\l aw where this paper was accomplished. We are grateful for the warm hospitality of the Institute of Theoretical Physics of the University of Wroc\l aw.


\bigskip
\noindent
Piret Kuusk\\
Institute of Physics, University of Tartu, \\
Riia 142, 51014 Tartu, Estonia\\
E-mail: piret@fi.tartu.ee
\par\medskip\noindent
Eugen Paal\\
Department of Mathematics, Tallinn University of Technology, \\
Ehitajate tee 5, 19086 Tallinn, Estonia\\
E-mail: eugen.paal@ttu.ee
\end{document}